\documentclass{elsart}

    \usepackage{latexsym,bm,amsmath,amssymb,amsfonts}
    \usepackage{epsfig,graphics,graphicx}
    \usepackage{makeidx}
    \usepackage{citesort}

\long\def\comment#1{ }

\newcommand{\beq}{\begin{eqnarray}}
\newcommand{\eeq}{\end{eqnarray}}
\newcommand{\be}{\vspace{-.4cm}\begin{eqnarray}}
\newcommand{\ee}{\vspace{-.5cm}\end{eqnarray}}

\newcommand{\tr}{{\rm tr}}

\newcommand{\BQA}{\begin{eqnarray}}
\newcommand{\EQA}{\end{eqnarray}}

\def\simge{\mathrel{%
   \rlap{\raise 0.511ex \hbox{$>$}}{\lower 0.511ex \hbox{$\sim$}}}}
\def\simle{\mathrel{
   \rlap{\raise 0.511ex \hbox{$<$}}{\lower 0.511ex \hbox{$\sim$}}}}

\newcommand{\x}{\bm x}
\newcommand{\y}{\bm y}

\newcommand{\z}{\bm z}

\begin{document}
\begin{flushright}
~\vspace{-1.25cm}\\
{\small\sf  RBRC-605}

\end{flushright}
\vspace{2.cm}

\begin{frontmatter}

\parbox[]{16.0cm}{ \begin{center}
\title{CGC formalism with two sources}

\author{Yoshitaka Hatta}

\address{ RIKEN BNL Research Center, Brookhaven National Laboratory,
Upton, NY 11973, USA}

\date{\today}
\vspace{0.8cm}
\begin{abstract}
In this work we extend the JIMWLK formalism to the two--source
problem. The $S$--matrix for the forward scattering can be written
in a double
 functional integral
 representation which involves
 the usual functional integral for the gluon field
  and  the spin path integral  for the external color sources.
  Modifications needed in
the light--cone gauge are discussed. Using our source term we
compute the produced gluon field and discuss the duality of the
high energy evolution kernel in the $pA$ collision.
\end{abstract}
\end{center}}

\end{frontmatter}

\section{Introduction}

Since the pioneering work \cite{larry}, impressive progress has
been made in the semiclassical description of high energy
scattering in QCD. It has been argued that, when accelerated to
very high energy, all hadrons and nuclei attain a universal form
of matter, the Color Glass Condensate (CGC). A hadron in the
infinite momentum frame is pictured as a collection of random,
classical charges $\rho$ which produce strong color fields $A^\mu
\sim 1/g$ (collective excitation of the small--$x$ partons). The
key issues of high energy scattering, such as the BFKL growth of
the gluon number \cite{bfkl} and its eventual saturation
\cite{levin}, are effectively encoded in the quantum evolution of
the CGC as summarized in the B--JIMWLK equation \cite{B,JIM,rg}.

In the JIMWLK formalism, color charges $\rho$ are treated as
\emph{static}, that is, independent of $x^+$ for a right--mover
(the `target'). The rationale behind this assumption is
well--known, and this is where the name ``glass" comes from:
Viewed from the natural time scale of the small--$x$ partons,
classical charges representing high--$x$ partons are `frozen' due
to Lorentz dilation. Implicit in this picture is an assumption
that the effect of the left--mover (the `projectile') is
negligible and one is effectively dealing with a single source
problem. Indeed, the original derivation of the JIMWLK equation
\cite{JIM,rg} makes no reference to the projectile.

However, in the recent surge of activity to go beyond the
B--JIMWLK equation (the \emph{Pomeron loop} physics), it has
become clear that  this initial picture must be modified at
important levels. In Ref.~\cite{HIMST} it has been recognized that
in order to include a new effect which is missing \cite{edmond} in
the JIMWLK equation, one has to abandon the
 assumption of $x^+$--independence. This effect, the \emph{gluon
Bremsstrahlung}, probes the multi--point correlation inside the
target in the $x^+$ direction, and therefore it forces  one to
consider $\rho$ as explicitly $x^+$--dependent variables. Another
observation, due to \cite{KL1}, is that in the dilute regime where
a hadron develops gluon number fluctuations, the
non--commutativity of color charges should be taken into account.
Again this is something that was neglected
 in the JIMWLK formalism from the outset for an obvious reason that
 commutators are suppressed $[\rho,\rho] \ll \rho\rho$ when a hadron
 is at saturation $\rho \sim 1/g^2$. \footnote{However, even when
 $\rho$ is large, quantum evolution of the CGC is actually sensitive to the commutator.
 See, Section~2.2 of \cite{hatta}.}

These two apparently different issues, the $x^+$--dependence of
$\rho$ and the non--commutativity, are nevertheless deeply related
through the process of the \emph{spin path integral} \cite{wz}. In
a previous publication \cite{hatta}, we have fully worked out this
connection in the single source problem. In this paper, we extend
this analysis to the case of the two--source problem, which we
propose as a model for the collision of two
 saturated hadrons (nuclei). In fact, there is already a CGC--based approach to the $AA$
 collision
  \cite{km,kras} in a special gauge which is useful under the
 assumption that the sources are strictly proportional to the delta function in the
 longitudinal direction.
 In prospect of applications to  quantum evolution,
 we relax this assumption and formulate the problem in general
 gauges. Our main goal is to set up a simple
 framework to study the  scattering process at \emph{any} stage of its high energy evolution,
 all the way from the gluon number fluctuation in the early stages\footnote{For a recent
discussion on the importance of fluctuations in the dilute regime,
see \cite{dio}.}
 to the quantum evolution of the $AA$ collision.

In Section 2, we derive the source term for the hadron--hadron
collision symmetrically for each of the hadrons starting from the
eikonal formula of the QCD $S$--matrix. The main ingredient is the
path integration of color matrices. The result, Eq.~(\ref{wei}),
has an obvious problem with the light--cone gauge, and we suggest
 an alternative source term specific to this gauge in Section 3.
 At the end of Section~3, we compare our approach
  with related works by Lipatov \cite{lipatov} and by Balitsky \cite{ian}.
 Sections 4 and 5 are applications.
 In Section~4, we outline how to compute the produced gluon field using our formulation.
  A known result for the $pA$ collision is reproduced.
  In Section~5, we discuss the
duality of the Bremsstrahlung Hamiltonian \cite{KL1,KL2,HIMST}.
 In Appendix~A, we give a brief review of the spin path integral
which is used throughout this paper.

\section{Eikonal scattering amplitude}
\setcounter{equation}{0}

 Let us start with high energy quark--quark scattering. In
 the eikonal approximation, the $S$--matrix for the forward scattering in the impact parameter
  space is given by
 a product of two Wilson lines \cite{nachtmann} \begin{align}
S(Y)=\langle V^{\dagger}(\x)W(\y) \rangle
 = \int [DA^\mu]\,  V^{\dagger}(\x)W(\y)e^{iS_{\text{YM}}[A]},
 \label{1}
 \end{align} where $Y$ is the rapidity and $\x-\y$ is the impact parameter. Letters in boldface denote
 two--dimensional transverse vectors. $V$ and $W$ are Wilson
 lines along the trajectories of colliding quarks \begin{align}
 V^\dagger(\x)=P\exp \left(ig\int^\infty_{-\infty} dx^- A^+_a(x^-,\x)t^a
 \right), \label{sau} \\ W(\x)=P\exp \left(ig\int^\infty_{-\infty} dx^+
 A^-_a(x^+,\x)t^a \right), \label{indi}\end{align}
 where $P$ is the  path--ordering operator and $t^a$ are the color matrices
 in the fundamental
 representation. In reality quarks have mass, so the trajectories
 are slightly tilted from the light--cone axes by a rapidity dependent
 angle. But for notational simplicity, we keep the deviation from the
 light--cone implicit. Also, gauge fixing (e.g., the Feynman gauge)
 is understood in Eq.~(\ref{1}) and
 we suppress the matrix indices
 of the Wilson lines.

In QED,  Eq.~(\ref{1}) can be exactly evaluated if one neglects
Fermion loops \cite{aba}. In QCD, on the other hand, it is a very
difficult problem and one needs some approximation schemes. One of
the popular approaches is the semiclassical method where one
reduces the problem to solving the classical equation of motion.
While the effectiveness  of such a method  is not immediately
obvious for the quark--quark scattering, it is arguably the most
realistic approach for the collision of two saturated
hadrons/nuclei. An obstacle to the classical description of
Eq.~(\ref{1}) is the non--commutative color matrices  $t^a$ in the
Wilson lines. However, one can convert these matrices into
c--numbers using the method of spin path integral \cite{wz}
\begin{align} W(\x)=\int [D\rho(x^+)]\,  e^{iS_{\text{WZ}}[\rho(x^+)]
     -ig\int dx^+ \rho^a(x^+)A^-_a(x^+,\x)}, \label{form} \end{align}
 where the color charges  $\rho(x^+)$ are commutative, but time dependent.
  $S_{\text{WZ}}[\rho]$ is a \emph{Wess--Zumino term}. In Apppendix, we
  give a quick
  derivation (and a more precise form) of Eq.~(\ref{form}).
      In the context of high energy scattering,
  the Wess--Zumino term was first introduced  in \cite{KL1}. Its origin was
   clarified later in \cite{hatta}.
 Using the formula Eq.~(\ref{form}) for both $W$ and $V$, one
 obtains an alternative expression of the $S$--matrix (the subscript L(R) is for the
  left--(right--)mover)
  \begin{align}
   & S_Y=\int [D\rho_{\text{R}}(x^+)D\rho_{\text{L}}(x^-)]\,  e^{iS_{\text{WZ}}[\rho_{\text{R}}(x^+)]+iS_{\text{WZ}}[\rho_{\text{L}}(x^-)]}
     \nonumber \\ &\qquad \qquad \times \int [DA^\mu]\,
     e^{iS_{\text{YM}}[A]
     -ig\int dx^- \rho_{\text{L}}(x^-)A^+_a(x^-,\y)
     -ig\int dx^+ \rho_{\text{R}}^a(x^+)A^-_a(x^+,\x)}. \label{spi}
 \end{align}
 In Eq.~(\ref{spi}), two independent functional integrals are involved;
   one for the gauge field and one for the color charges.
  Thanks to the latter, the external
 non--commutative charges are treated as the usual c--number source term $\int dx
  J^\mu A_\mu$ as in the QCD generating functional.

We now generalize the quark--quark scattering formula
Eq.~(\ref{spi}) to  scattering of more complicated objects.
Following the original suggestion \cite{larry}, we model a hadron
or a nucleus at very high energy as a bunch of partons belonging
to various representations of the color group and continuously
distributed in the transverse plane as well as in the longitudinal
 (i.e., $x^-$ for the right--mover) direction.
  For a thick nucleus, or a highly evolved
 hadron, the representations can be very large. Each parton propagates
eikonally and leaves a Wilson line in the corresponding
representation in its wake. Thus we propose the following
representation of the $S$--matrix
\begin{align} & S_Y=\langle  \prod_i V^\dagger(\x_i) \prod_j
W(\y_j) \rangle \nonumber \\ & \qquad =\int
[D\rho_{\text{L}}(x^-)D\rho_{\text{R}}(x^+)]Z[\rho_{\text{L}}]Z[\rho_{\text{R}}]
\, e^{iS_{\text{WZ}}[\rho_{\text{R}}(x^+)]+iS_{\text{WZ}}[\rho_{\text{L}}(x^-)]} \nonumber \\
& \qquad \ \ \quad \times \int [DA^\mu]\, e^{iS_{\text{YM}}[A]
-ig\int d^4x \rho_{\text{L}}A^+_a
     -ig\int d^4x \rho_{\text{R}}^aA^-_a}. \label{wei}
\end{align} Note that integrals in the source terms are four--dimensional.
 Viewed from a distance, the distribution of charges is localized in the
longitudinal  direction and may be regarded as a delta function in
the first approximation
$\rho_{\text{R}}(x)\sim\delta(x^-)\rho_{\text{R}}(x^+,\x)$ and
$\rho_{\text{L}}(x)\sim \delta(x^+)\rho_{\text{L}}(x^-,\x)$,
though it is known that regularization is needed for certain
applications. For a `dilute' hadron which has not yet gone through
high energy evolution, $\rho$ is of order 1, whereas for a `dense'
hadron, $\rho$ can be as large as $\sim 1/g^2$ characteristic of
 saturation.
$Z[\rho]$ is a weight function which describes the probability
 distribution of partons at $x^\pm \to -\infty$ like the one in the dipole
model \cite{al}.

In practice one is typically interested in the scattering of
globally color neutral  objects. In this case it is possible to
recast Eq.~(\ref{wei}) in a gauge invariant form. One can
construct a closed loop out of a pair of Wilson lines
corresponding to a charge and an anti--charge at different
transverse points by attaching transverse Wilson lines $U$ at
$x^-=\pm \infty$
\begin{align} V^\dagger(\x_i)V(\x_j) \to
U(\x_j,\x_i,x^-=\infty)V^\dagger(\x_i)U(\x_i,\x_j,x^-=-\infty)V(\x_j).
\label{napo}
\end{align} The path connecting $\x_i$ and $\x_j$ is arbitrary, and we choose it to extend
to spatial infinity and then come back \begin{align}
U(\x_j,\x_i)=U^\dagger({\bm \infty},\x_j)U({\bm \infty},{\bm
\infty}')U({\bm \infty}',\x_i)
\end{align} This procedure is repeated for all existing color
singlet units (dipoles, quadrupoles, etc.) in the projectile, and
similarly for the target. Then the $S$--matrix Eq.~(\ref{wei})
acquires a gauge invariant meaning. Indeed, since the field
strength $F_{ij}$ vanishes at infinity $x^\pm =\pm \infty$, it is
possible to gauge away transverse fields $A_i(x^\pm=\pm
\infty,\x)$. With this gauge choice a Wilson loop reduce to a
product of light--like Wilson lines as in Eq.~(\ref{wei}).
However, this also means that in generic gauges Eq.~(\ref{wei})
should be understood that at each end point of Wilson lines a
transverse Wilson line extending to infinity $U({\bm \infty},\x)$
is attached.\footnote{Incidentally, we note that for a closed loop
it is possible to rewrite the Wess--Zumino term in a more elegant
way than that given in the appendix. In the case of color SU(2),
\begin{align} P\exp\left( ig\oint dx^\mu A_\mu^a t^a
\right)=\int [D\rho(t,s)]\exp\left(-ig\oint dt \frac{dx^\mu}{dt}
A_\mu^a \rho^a(t) + iJ\oint dt \int_0^1 ds (\partial_t
\vec{\rho}\times
\partial_s \vec{\rho})\cdot \vec{\rho} \right),
\end{align} where $t$ is the coordinate along the contour (the `ordering variable') and $s$ is the
radial coordinate of the disc spanned by the contour. $\rho(t,s)$
is a continuation of $\rho(t)$ inside the disc such that
 at the boundary, $\rho(t,s=1)=\rho(t)$. This $(1+1)$--dimensional Wess--Zumino term
(often called the Polyakov spin factor) is the form suggested in
\cite{KL1}. }

Instead of using the path integral formula Eq.~(\ref{form}), one
could trivially rewrite the Wilson lines in the first line of
Eq.~(\ref{wei}) as
\begin{align} W(\x)=\int [D\rho_{\text{R}}(x^+)]\, e^{-ig\int dx^+
\rho_{\text{R}}(x^+)A^-(x^+,\x)} P\exp \left(\int^\infty_{-\infty}
dx^+ \frac{\delta}{\delta \rho_{\text{R}}^a(x^+)}t^a \right)
\delta [\rho_{\text{R}}(x^+)], \label{mea}
\end{align}
 and similarly for $V^\dagger(\x)$. This is in accord with the suggestion in
 \cite{KL5}
  that the weight function must contain Wilson lines of
 the functional derivative, $\delta/\delta \rho$,
  to ensure non--commutativity.
  We note that the derivative Wilson line turned out to be useful for the
    generalized dipole model \cite{KL4,dipole,marquet} (and in other context
 \cite{KL6}).
 Both Eq.~(\ref{form}) and Eq.~(\ref{mea}) are  exact rewritings of
 a Wilson line. However,  the meaning of the measure $[D\rho]$
 is completely different in the two cases. Eq.~(\ref{form}) is
 the genuine path integral of the `spin operator'  $t^a \to \rho^a$ and therefore
 the measure $[D\rho]$ respects the compactness of the underlying group (see Appendix).
 One can speak of
 its saddle point, gauge transformation, etc.
 [The saddle point approximation will be important
  in Section~4.] On the other hand, $\rho_{\text{R}}(x^+)$ in Eq.~(\ref{mea})
 is a dummy variable which comes from
 the  insertion of unity $1=\int [D\rho]\,  e^{-i\rho
 A}\delta[\rho]$.

\section{Light--cone gauge}
\setcounter{equation}{0} For the description of high energy
scattering, it is often very convenient to work in the light--cone
gauges, $A^+=0$ or $A^-=0$. The source terms in these gauges
assume a peculiar form and deserve a separate study. For
definiteness, we choose $A^+=0$. In this gauge, $V^{(\dagger)}=1$,
and as one can see from  Eq.~(\ref{napo}) the transverse Wilson
lines represent the dominant interaction with the left--mover.

 Instead of setting $A^+=0$ in Eq.~(\ref{napo}) from the beginning,
 let us reach this gauge via gauge rotation from
  other gauges in which $A_i(x^-=\pm\infty)=0$, that is, Eq.~(\ref{wei}) is valid
  without insertions of transverse Wilson lines.
Such a detour is useful since the relevant gauge rotation often
appears in the literature \cite{rg,HIMST}. Besides, it prescribes
 the boundary condition of the transverse field at $x^- \to \pm \infty$.
  The $A^+=0$ gauge can be achieved by the following gauge
 rotation \begin{align} A^\mu \to \tilde{A}^\mu= VA^\mu
 V^\dagger +\frac{i}{g}V\partial^\mu V^\dagger, \label{xi} \end{align} where
  (c.f., Eq.~(\ref{sau})) \begin{align}
 V^\dagger(x)=P\exp\left( ig\int^{x^-}_{-\infty} dz^- A^+_a(x^+,z^-,\x)t^a
 \right). \label{qqqqq} \end{align}  Then the time--like Wilson
line $V^\dagger$  in the original gauge can be rewritten  as  a
transverse Wilson line in the $\tilde{A}^+=0$ gauge
\begin{align} V^\dagger(\x)=P\exp\left( ig \int_{\x}^{\bm \infty}
dz^i \tilde{A}_i^a(x^-=\infty,\z) t^a \right) \equiv e^{-ig\phi^a
t^a}, \label{jik}
\end{align} where the integration path is arbitrary since
$\tilde{A}_i$ is pure gauge. To make contact with the light--cone
gauge perturbation theory, we note that to leading order in $g$,
$\tilde{A}_i \approx \partial_i \phi$, and Eq.~(\ref{jik}) can be
written as
\begin{align} V^\dagger(\x)\approx \exp \left(igt^a \frac{\partial^i}{\bm
\nabla^2}\tilde{A}_i^a(x^-=\infty)\right). \label{gene}
\end{align} The coupling
$\sim (\partial^i/{\bm \nabla}^2) \tilde{A}_i$ is precisely the
structure one finds from the direct perturbative analysis in the
light--cone gauge (see, e.g., Sect.~3.2 of \cite{yuan}).
Eq.~(\ref{gene}), or its continuum generalization $t^a \to -\int
dx^+d\x \hat{\rho}_L^a$ ($\hat{\rho}$ being non--commutative) is
the source term that is useful for our purposes.  Due to the long
range nature of the light--cone gauge propagator $\sim 1/p^+$,
sources at infinity $x^-=\infty$ cannot be neglected. The
scattering in this
 gauge is pictured as a left--moving quark  changing directions at $x^-=\infty$
 to the transverse direction,
 and the interaction with this
 quark entirely comes from this part of the propagation.

The usual assumption of saturation amounts to postulating  that
$\hat{\rho}_L \sim 1/g^2$ and dropping  non--commutativity by hand
$\hat{\rho}_L \to \rho_L$.\footnote{Note that the loss of
non--commutativity also justifies the approximation $\partial_i
\phi \approx \tilde{A}_i$.} In place of Eq.~(\ref{wei}), we now
have (omitting the tilde on $A^\mu$)
\begin{align}& \int
[D\rho_{\text{R}}(x^+)D\rho_{\text{L}}]Z[\rho_{\text{R}}]Z[\rho_{\text{L}}]
\, e^{iS_{\text{WZ}}[\rho_{\text{R}}(x^+)]} \nonumber \\
& \qquad \ \  \times \int_{A^+=0} [DA^\mu]\,  \exp
\left(iS_{\text{YM}}[A]
     -ig\int d^4x \rho_{\text{R}}^aA^-_a -ig\int_{x^-=\infty} dx^+d\x
 \rho_{\text{L}}^a \frac{\partial^i}{{\bm \nabla}^2}A_i^a \right). \label{lc}
\end{align}

The role of the source term $\sim \rho_L$ is to generate a
 classical field of the left--mover (see, \cite{ian} for a
 construction of source terms from this perspective)
\begin{align} A^-_{\mbox{\scriptsize{cl}}}=-\frac{g}{{\bm \nabla}^2}
\rho_{\text{L}}(x^+,\x).
 \label{apo} \end{align}
 Indeed, it precisely cancels the boundary term of the Yang--Mills
 action when expanded around Eq.~(\ref{apo})
\begin{align}& S_{\text{YM}}[A^-_{\mbox{\scriptsize{cl}}}+A^\mu]=
S_{\text{YM}}[A^-_{\mbox{\scriptsize{cl}}}]-\int [D_\nu^{cl}
A_\mu]
 F^{\nu \mu}_{\mbox{\scriptsize{cl}}} + {\mathcal O}(A^2)\nonumber \\
 & \qquad = \int [D_\nu^{cl} F^{\nu \mu }_{\mbox{\scriptsize{cl}}}]A_\mu -\int dx \partial_\nu
 (F^{\nu \mu}_{\mbox{\scriptsize{cl}}}A_\mu)+\cdots \nonumber \\
 & \qquad \sim \int [D_\nu^{cl} F^{\nu - }_{\mbox{\scriptsize{cl}}}]A^+ -\int dx
 \partial_-
 (F^{- i}_{\mbox{\scriptsize{cl}}}A_i)\nonumber \\
  & \qquad = \int dx^+d\x
 \partial^iA^{a-}_{\mbox{\scriptsize{cl}}}\bigl(A_i^a(x^-=\infty)-A_i^a(x^-=-\infty)\bigr)
 \nonumber \\ &\qquad =g\int dx^+d\x
 \rho_{\text{L}}^a \frac{\partial^i}{{\bm \nabla}^2}\bigl(A_i^a(x^-=\infty)
 -A_i^a(x^-=-\infty)\bigr). \label{qq}
 \end{align}

 In general, by choosing a prescription
 for the singularity $1/p^+$ of the light--cone gauge propagator, or equivalently,
  by fixing the residual gauge freedom of the $A^+=0$ gauge, one can require either
  $A^i(x^-=\infty)=0$
 or $A^i(x^-=-\infty)=0$ \cite{KA,rg,yuan}. Eq.~(\ref{lc}) corresponds to the latter
  (the final state interaction) in accordance with  the choice Eq.~(\ref{qqqqq}).
 Of course, one could have chosen the other boundary
 condition $A_i(x^-=\infty)=0$ (the initial state interaction) by changing $-\infty \to \infty$ in Eq.~(\ref{qqqqq}).

The above procedure is valid only when the effect of the
right--mover is negligible, $\rho_{\text{R}}\sim 1$. If the
right--mover is at saturation, $\rho_{\text{R}} \sim 1/g^2$, it
would create a strong classical background field $\sim 1/g$ and
the source term must be modified accordingly. We assume that, in
the original gauge, \begin{align} A^+=\alpha+\delta A^+,
\end{align} where $\alpha \sim 1/g$ is the classical field of the right--mover.
We then transform to the $A^+=0$ gauge in two steps. First we
rotate away the classical component \begin{align} A^\mu \to
\tilde{A}^\mu= V_0A^\mu
 V^\dagger_0 +\frac{i}{g}V_0\partial^\mu V_0^\dagger, \label{xi}
 \end{align} where \begin{align} V_0^\dagger(x)=P\exp\left( ig
 \int^{x^-}_{-\infty}
dz^- \alpha(z^-) t^a \right)\equiv e^{-ig\chi}. \end{align} In
this gauge,
\begin{align} \tilde{A}^+=V_0\delta A^+ V_0^\dagger, \qquad \tilde{A}^i(x^-=\infty)
=\frac{i}{g}V_0\partial^i V_0^\dagger \equiv B^i. \end{align} Then
we remove the remaining weak field $\tilde{A}^+$ by the gauge
rotation with
\begin{align} \delta V^\dagger=P\exp\left( ig
 \int^{x^-}_{-\infty}
dz^- \tilde{A}^+(z^-) t^a \right) \equiv e^{-ig\phi}. \end{align}
 The transverse field at infinity becomes
\begin{align}
 \tilde{A}_i(x^-=\infty) \to \delta VB_i\delta V^\dagger +\frac{i}{g}\delta V
 \partial_i\delta V^\dagger
 \approx
 B_i + D_i \phi, \label{boun} \end{align} where $D_i \equiv \partial_i-igB_i$.
 Assuming that the left--mover is also saturated, we can let $t^a_{\text{L}}\to -\int
 dx^+d\x \rho^a_{\text{L}}$ and express
 $V^\dagger$ in the original gauge in terms of the boundary transverse field
 in the light--cone gauge Eq.~(\ref{boun}).
 \begin{eqnarray} V^\dagger=V^\dagger_0\delta V^\dagger=e^{-ig\chi}e^{-ig\phi}
 &\to& e^{ig\int \rho^a_{\text{L}}\chi^a}e^{ig\int \rho^a_{\text{L}}\phi^a}\nonumber \\
  &=&e^{i\int D_iD^iA^-_{\text{cl}}\chi+i\int D_iD^iA^-_{\text{cl}}\phi}\nonumber \\
  &=&e^{-i\int D^iA^-_{\text{cl}}D_i\chi-i\int D^iA^-_{\text{cl}}D_i\phi}
 \nonumber \\ &\approx& \exp\left( i\int_{x^-=\infty} dx^+d\x
 F^{-i}_{\text{cl}}(B_i+D_i\phi) \right), \label{31} \end{eqnarray} where $F^{-i}_{\text{cl}}$ is
 the field strength constructed from the two classical background
 fields $(B_i,\ A^-_{\text{cl}})$ with $D_\nu F^{\nu -}_{\text{cl}}=
 D_iD^iA^-_{\text{cl}}=g \rho_{\text{L}}$ (c.f., Eq.~(\ref{apo})).
 In the last line, we have used $\partial_i \chi \approx B_i$ and neglected the commutator $[B_i, \chi]
 \sim [B_i(\x), \int_{\x} dz^i B_i(\y)]$ assuming that $B_i$ does not change much
 over the typical separation of a charge and an anti--charge in a dense projectile.
  Eq.~(\ref{31}) is the desired source term for the left--mover with the field of
  the right--mover taken into account. We see from  Eq.~(\ref{qq}) that its linear
  part precisely
  cancels the  boundary
term  of the Yang--Mills action when expanded around the
background field $A^\mu \to B^i+A^-_{\text{cl}}+A^\mu$. (In fact,
it also cancels $S_{\text{YM}}[A^\mu_{\text{cl}}]$. See,
Eq.~(\ref{hen}) below.)

\vspace{8mm}

Summarizing, Eq.~(\ref{wei}) and Eqs.~(\ref{lc}), (\ref{31})  are
our setup for the two--source problem.  In the semiclassical
 framework one first solves  the  Yang--Mills equation with sources obtained by
 varying with $A^\mu$ in Eq.~(\ref{wei})
\begin{align} D_\nu F^{\nu \mu}=\delta^{\mu +}
\rho_{\text{R}}(x^+) + \delta^{\mu -}\rho_{\text{L}}(x^-), \qquad
D^-\rho_{\text{R}} + D^+\rho_{\text{L}}=0. \label{cla}
\end{align}  The second equation, the current conservation law,
  is a consequence of the first.
  Although Eq.~(\ref{cla}) is frequently written down in the
literature of the CGC, the precise meaning of the time dependence
of charges, or from which source term the equation has been
derived is not always articulated. This latter point deserves
further comments. Assuming that the currents are separately
conserved,
\begin{align} D^+_{\mbox{\scriptsize{cl}}}\rho_{\text{L}}=0,
\qquad D^-_{\mbox{\scriptsize{cl}}}\rho_{\text{R}}=0, \label{se}
\end{align}
 in our approach we can  integrate these equations with an arbitrary boundary condition,
  in particular, the \emph{retarded } boundary
 condition \begin{align} \label{retarded}
 \rho_{\text{L}}(x^-)=V^\dagger(x^-)\rho_{\text{L}}(-\infty)V(x^-),
 \nonumber \\
 \rho_{\text{R}}(x^+)=W(x^+)\rho_{\text{R}}(-\infty)W^\dagger(x^+),\end{align}
 which is usually employed for the gluon production problem.
 [$V(x^-)$ ($W(x^+)$) is the same as in Eq.~(\ref{sau})
(Eq.~(\ref{indi})) except that the upper limit of the integration
is $x^-$
 ($x^+$).] On the other hand,  currents derived from effective source
 terms of the JIMWLK formalism such as $S_{\text{W}}=\int d^3x \tr
 [\rho_{\text{R}} \widetilde{W}]$ and $S_{\text{W}}=\int d^3x
 \tr[\rho_{\text{R}}\ln \widetilde{W}]$ ($\widetilde{W}$ being the adjoint Wilson
 line)
  \cite{JIM,rg,raju,fuku} via $J^+=\delta S_{\text{W}}/\delta
 A^-$ are covariantly conserved, $D^-J^+=0$, but one cannot freely choose the boundary
 condition to this differential equation because it is already prescribed in
 the structure of $S_{\text{W}}$. In fact, it has been pointed out in \cite{rg}
 that \emph{any} effective source term of the form
 $S_{\text{W}}[\rho,W]$ is incompatible with the retarded boundary
 condition. Our source term with the Wess--Zumino term is not subject to such a constraint.

In practice, one has to resort to some sort of perturbative
expansion to solve Eq.~(\ref{cla}). The Yang--Mills action plus
source terms evaluated at this saddle point is often referred to
as the \emph{effective action}. Clearly, the choice of the source
term and how one expands the saddle point solution are the key
issues. We shall work out the case of the $pA$ collision in
Section~5
 and leave the analysis of the $AA$ collision to future works.
[This was partly done in \cite{HIMST}.] In the reminder of this
section we draw a comparison (albeit superficially)  between our
 approach and two previous approaches in the literature. Our main
 concern here is the form of the source term.

 (i) In the context of quark--quark
scattering, Lipatov \cite{lipatov} constructed an effective action
describing the interaction between the $s$--channel `physical'
gluons $V^\mu$ and the $t$--channel `reggeized' gluons $A^{\pm}$
in a given rapidity interval
\begin{align} S_{\text{eff}}[v]= S_{\text{YM}}[v] -\frac{2i}{g}\int dx^-d\x
\, \tr[W[v^-]\bm{\nabla}^2A^+] -\frac{2i}{g} \int dx^+d\x \,
\tr[V^\dagger[v^+] \bm{\nabla}^2A^-]+\cdots, \label{lev}
\end{align}
 where $v^\mu(x)=A^+(x^-,\x) +A^-(x^+,\x) +V^\mu(x)$ is the total
 gauge field.
 Eq.~(\ref{lev}) describes quasi--elastic
 processes (linear terms in $A^\pm$),
  the production of gluon clusters in the multiregge kinematics
 (quadratic terms in $A^\pm$), and so on.  One can integrate out the $V^\mu$ field by
  solving the classical equation of motion
 $\delta S_{\text{eff}}/\delta v^\mu=0$ perturbatively in powers of $g$.
  The effective action evaluated at this solution
  $S_{\text{eff}}[v_{\text{cl}}[A^\pm]]$  contains all possible
 reggeon number changing vertices in the $t$--channel
 $(A^+)^m \to (A^-)^n$ in a given rapidity interval.
  Clearly, there is a close correspondence with the effective action for the
  two--source problem
  in the CGC approach. One notices that
   the structure\footnote{ Wilson lines in Eq.~(\ref{lev})
   arise after determining the boundary
   condition for the inverse operator $\partial^{-1}_{\pm}$ in the multi--gluon production
   vertices.
   A more fundamental expression
    of the interaction term can be found in \cite{lipatov}.  } of the interaction terms in Eq.~(\ref{lev})
    is identical to the source term of the JIMWLK formalism
 $S_{\text{W}}\propto \int d^3x \tr
 [\rho_{\text{R}} \widetilde{W}]$ \cite{JIM,rg}. It is thus tempting to identify
  the reggeon fields in \cite{lipatov} with the classical color charges of
  \cite{JIM,rg}
  \begin{align} \bm{\nabla}^2 A^+ \leftrightarrow g\rho_{\text{R}}(x^-,\x),
 \nonumber \\ \bm{\nabla}^2 A^- \leftrightarrow g\rho_{\text{L}}(x^+,\x).
 \label{lip} \end{align}   Eq.~(\ref{lip}) may be related to the technicality
 in \cite{lipatov} that the reggeon fields
 are invariant under local gauge transformations. It would be important
 to study how the effective action in the Wilson line representation
  \cite{ian,HIMST} is matched
 with the perturabative expansion in \cite{lipatov}.

(ii) Instead of introducing color charges, Balitsky described the
colliding hadrons as  shock waves, i.e., fixed gauge field
configurations \cite{ian}. Eq.~(\ref{wei}) may be compared with
the twice rapidity--factorized formula for the scattering
amplitude
\begin{align} & iT_Y(p_a,p_b \to p'_a,p'_b) =\int [ DB^\mu
DC^\mu]\, e^{iS_{\text{YM}}[B]+iS_{\text{YM}}[C]}
J(p_a)J(p_b)J(p'_a)J(p'_b) \nonumber \\ & \qquad \qquad \qquad
\qquad \qquad \times \int [DA^\mu]\,  e^{iS_{\text{YM}}[A]+i\int
 d\x
f^i_a[B] f_i^a[A] +i\int
 d\x
g^i_a[A] g_i^a[C] }, \label{ni} \end{align}
 where $J$'s are the currents of external hadrons and
 \begin{align} & f_i[A]=\int_{-\infty}^{\infty}
dx^+W^\dagger(x^+)F_{+i}(x^+)W(x^+), \nonumber
\\& g_i[A]=\int_{-\infty}^{\infty}
dx^-V(x^-)F_{-i}(x^-)V^\dagger(x^-). \label{line}
\end{align}  The source terms in Eq.~(\ref{ni}) perfectly makes sense
  in any gauge including the
 light--cone gauges.
  The gauge fields $A,B,C$ are ordered in rapidity,
 $Y_{B}>Y_1>Y_{A}>Y_2>Y_{C}$ and the Wilson lines in $f$ and $g$
 are actually not strictly light--like but along the direction of the
 rapidity divider $Y_1$ and  $Y_2$, respectively. Viewed from
 the $A^\mu$ field, Eq.~(\ref{ni}) is a two--source problem
 just like Eq.~(\ref{wei}). It has been suggested \cite{ian}
 that the classical Yang--Mills
 equation with sources
 $\sim \delta f_i/\delta A^- $, $\delta g_i/\delta A^+$ can be formally solved
 via the commutator expansion.

\section{Gluon production}
\setcounter{equation}{0}

 Though Eqs.~(\ref{wei}), (\ref{lc}) were derived as the $S$--matrix for
 the elastic scattering,
 proper specification of the source terms should allow one to compute
 the particle production
  in hadron collisions as already implied around
  Eq.~(\ref{retarded}).
 In this section we demonstrate that this is indeed the case. The produced gluon field
 (the inclusive gluon production)
  can be computed via straightforward manipulations
  of the functional average of the gauge field in the presence of sources (Hereafter we often omit `$\int
   d^4x$' to simplify the appearance of equations. )
 \begin{align}& \langle A^\mu(x) \rangle= \int
 [D\rho_{\text{L}}(x^-)
 D\rho_{\text{R}}(x^+)]\,  e^{iS_{\text{WZ}}[\rho_{\text{R}}(x^+)] +
 iS_{\text{WZ}}[\rho_{\text{L}}(x^-)]} \nonumber \\ &
 \qquad \qquad  \qquad \times \int [DA^\mu]\, e^{iS_{\text{YM}}[A]-i\rho_{\text{R}} A^- -i\rho_{\text{L}} A^+ } A^\mu(x),
\label{www} \end{align}  with the premise that the last propagator
contracted with $A^\mu(x)$ should be the retarded one. Typically
one is interested in the behavior in the forward light--cone
$x^\pm >0$. To see how Eq.~(\ref{www}) works in practice, below we
consider two examples, the $pp$ and the $pA$ collisions. In both
 cases we shall illustrate the nontrivial  roles played by the
Wess--Zumino term.

\subsection{The $pp$ collision}
 When both charges are small $\rho_{\text{R}/\text{L}}\sim 1$ (the
$pp$ collision),
 $\langle A^\mu(x) \rangle$ can be computed in perturbation theory by
expanding $g$ dependent terms (source terms as well as the
interaction terms in the Yang--Mills action). One of the
contributions to $g^3$ is given by
\begin{align}  \frac{(-ig)^3}{2} \langle \rho_{\text{R}}A^-(w) \rho_{\text{R}}A^-(y) \rho_{\text{L}} A^+(z) A^\mu(x)
\rangle = (-ig)^3G^{\mu -}_0(x,w)G^{-+}_0(y,z) \langle
\rho^a_{\text{R}}(w)\rho^b_{\text{R}}(y)\rho^a_{\text{L}}(z)\rangle,
\end{align}
 where $G_0^{\mu \nu}$ is the free gluon propagator. The product of
 charges should be handled a little carefully. Due to the presence of
 the Wess--Zumino term, the following identity holds (c.f., Eq.~(\ref{abo})) \begin{align}
 \langle \rho_{\text{R}}(w^+,\bm{w}) \rho_{\text{R}}(y^+,\y)\rangle=
 \theta (w^+ - y^+)\hat{\rho}_{\text{R}}^a(\bm{w})\hat{\rho}_{\text{R}}^b(\y)+\theta (y^+ - w^+)
 \hat{\rho}_{\text{R}}^b(\y)\hat{\rho}_{\text{R}}^a(\bm{w}), \label{pro}
 \end{align} where $\hat{\rho}_{\text{R}}$ is the non--commutative, but
 $x^+$--independent charge of the right--moving proton
 \begin{align}
\hat{\rho}_{\text{R}}^a(\bm{w})\hat{\rho}_{\text{R}}^b(\y)-\hat{\rho}_{\text{R}}^b(\y)\hat{\rho}_{\text{R}}^a(\bm{w})
=-if^{abc}\hat{\rho}^c_{\text{R}}(\y)\delta(\bm{w}-\y).
\end{align} ($\hat{\rho}_{\text{L}}$ and $\hat{\rho}_{\text{R}}$ commute.)
 We decompose the product in Eq.~(\ref{pro}) into symmetric and asymmetric
 parts to get
  \begin{align} \frac{-i}{2}f^{abc}\bigl(\theta(w^+ -y^+)-\theta
(y^+ -w^+)\bigr)\hat{\rho}_{\text{R}}^c(\y)\delta(\bm{w}-\y)+
\frac{1}{2}\{\hat{\rho}^a_{\text{R}}(\bm{w}),\hat{\rho}^b_{\text{R}}(\y)\}.
\label{ver}
\end{align}
 The symmetric part does not contribute when we put the detected
 gluon on--shell, so it can be discarded. Other terms are similarly
 computed and the result is identical to \cite{kovchegov2}.\footnote{ Except for the
  $\epsilon$--prescription of internal Fermion propagators
  $\theta(w^+ -y^+) \leftrightarrow \frac{1}{2}(\theta(w^+ -y^+)-\theta(y^+ -w^+))$
   which, however,
  does not affect the result if the outgoing gluon is put
  on--shell.}

\subsection{The $pA$ collision}

 Next we consider the $pA$ collision with the
 proton moving to the right. By this we mean that
 $\rho_{\text{R}}
\sim 1$ and $\rho_{\text{L}} \sim 1/g^2$. This problem has been
 extensively studied in the literature \cite{KA,glu,ja,tan}, so we merely intend to show how things work
in our formulation.
 It is convenient to choose the $A^+=0$ gauge.
   The produced gluon field is (see Eq.(\ref{lc}))
\begin{align} \langle A^\mu (x)\rangle= \int_{A^+=0}
[D\rho_{\text{R}}(x^+)] \int [DA^\mu] \,
e^{iS_{\text{YM}}[A]-ig\rho_{\text{R}} A^-
+ig\frac{\partial^i}{\bm{\nabla}^2}
\rho_{\text{L}}A_i(x^+,\infty,\x)+iS_{\text{WZ}}[\rho_{\text{R}}(x^+)]
}A^\mu(x).
\nonumber \\
  \label{mi} \end{align}
  Unlike in the $pp$ collision, one cannot expand in powers of $\rho_{\text{L}} A^i$
   since $\rho_{\text{L}}$ is large.
  However, this term can be removed (see Eq.~(\ref{qq})) by making the following shift
  \begin{align} A^\mu =A^-_{\mbox{\scriptsize{cl}}}+a^\mu, \end{align}
 with $A^-_{\mbox{\scriptsize{cl}}}$ given in Eq.~(\ref{apo}).
 Then Eq.~(\ref{mi}) becomes
  \begin{align} \langle A^\mu (x)\rangle=\int
  [D\rho_{\text{R}}(x^+)]\,
e^{iS_{\text{WZ}}[\rho_{\text{R}}(x^+)] }\int_{a^+=0} [Da^\mu] \,
e^{ia^\mu D^{-1}_{\mu \nu}[A^-_{\mbox{\scriptsize{cl}}}]a^\nu
-ig\rho_{\text{R}} (A^-_{\mbox{\scriptsize{cl}}} +a^-)
}(A^-_{\mbox{\scriptsize{cl}}}+a^\mu). \nonumber
\\ \end{align}   Let us focus on the transverse
 field $\langle A^i (x)\rangle$, since this is the only component which contribute
 to the gluon production in this gauge.
  Since $\rho_{\text{R}}$ is small, we expand only to first order in
  $\rho_{\text{R}} a^-$.
\begin{align} \langle A^i(x) \rangle =\int [D\rho_{\text{R}}(x^+)]\, e^{
iS_{\text{WZ}}[\rho_{\text{R}}(x^+)]}\int [Da^\mu]\,  e^{ia^\mu
D^{-1}_{\mu \nu}[A^-_{\mbox{\scriptsize{cl}}}]a^\nu
-ig\rho_{\text{R}} A^-_{\text{cl}} } (-ig)\int_y \rho_{\text{R}}
a^-(y)a^i(x) \nonumber
\\ = -ig\int [D\rho_{\text{R}}(x^+)]\,  e^{
iS_{\text{WZ}}[\rho_{\text{R}}(x^+)]-ig\rho_{\text{R}}
A^-_{\text{cl}}}\int_{y^-=0} dy^+d\y
G^{i -}(x,y)[A^-_{\text{cl}}]\rho_{\text{R}}(y^+,\y), \nonumber \\
\label{naotan}
\end{align} where $G^{\mu \nu}[A^-_{\mbox{\scriptsize{cl}}}]$ is the gluon propagator
 in
the $a^+=0$ gauge in the presence of the background field
$A^-_{\mbox{\scriptsize{cl}}}$. In Eq.~(\ref{naotan}), we can
safely set
$\rho_{\text{R}}(y^+,y^-,\y)=\rho_{\text{R}}(y^+,\y)\delta(y^-)$.
   It is possible to do the saddle point approximation for the path integral
   $D\rho_{\text{R}}(x^+)$  since
    a large factor $A^-_{\mbox{\scriptsize{cl}}}$ multiplies $\rho_{\text{R}}$. The saddle point equation
    is nothing but the current conservation law \cite{hatta}
    \begin{align} & D_{\text{cl}}^-\rho_{\text{R}}(x^+)=0, \nonumber \\
   & \to \rho_{\text{R}}(y^+,\y)=W(y^+,\y)\rho_{\text{R}}(\y)
   =P\exp\left(ig\int^{y^+}_{-\infty}dz^+
   A^-_{\text{cl}}(z^+,\y)\right)\rho_{\text{R}}(\y). \label{nao}
   \end{align}
   At the saddle point,
   $S_{\text{WZ}}[\rho_{\text{R}}(x^+)]-g\rho_{\text{R}}
A^-_{\mbox{\scriptsize{cl}}}=0$ \cite{hatta}.
 The background field propagator
 can be extracted from \cite{larry,rg}   \begin{align} &
G^{i-}(x,y)=\int
 dp^+\frac{-ie^{-ip^+(x^--y^-)}}{p^+}\Bigl(G_0(\vec{x}-\vec{y},p^+)\theta(y^+)
 \partial_y^{\dagger i}
 \nonumber \\ & \qquad \qquad \qquad +2ip^+\int d\vec{z}
 G_0(\vec{x}-\vec{z},p^+)\delta(z^+)G_0(\vec{z}-\vec{y},p^+)\theta(-y^+)W_z
  \Bigr) \partial_y^{\dagger i} \nonumber \\ & \qquad \qquad=
  2\pi \int dz^- \theta(z^-)G_0(\vec{x}-\vec{y},x^--y^--z^-)\theta(y^+)
 \partial_y^{\dagger i} \nonumber \\ &  \qquad
 \qquad \qquad +4\pi \int dz
 G_0(x-z)\delta(z^-)G_0(z-y)\theta(-y^+)W_z
   \partial_y^{\dagger i}, \label{sig}
 \end{align}
  where $\vec{x}=(x^+,\x)$ and \begin{align} W_y= P\exp
  \left(ig\int^\infty_{-\infty} dy^+ A^-_{\text{cl}}(y^+,\y)
 \right). \end{align}  In Eq.~(\ref{sig}), we have dropped terms which do not contribute
 at $x^+ >0$.
 $\partial^\dagger$ means that the derivative acts on functions on its left.
  $G_0$ is the  retarded free scalar propagator.
   In the second equality, we have chosen the light--cone gauge prescription
    $1/p^+ \to 1/(p^+ +i\epsilon)$ to go with the boundary condition we have chosen.
 After the substitution of Eqs.~(\ref{nao}) and (\ref{sig}), Eq.~(\ref{naotan})
 takes the form
 \begin{align}
&\langle A^i(x) \rangle =-2\pi ig\int_{y^-=0} dy^+d\y
 \Bigl( \int dz^- \theta(z^-)G_0(\vec{x}-\vec{y},x^- - z^-)\theta(y^+)
 \partial_y^{\dagger i} \nonumber \\ & \qquad \qquad \qquad +2 \int dz
 G_0(x-z)\delta(z^+)G_0(z-y)\theta(-y^+)W_z
   \partial_y^{\dagger i} \Bigr) W(y^+,\y)\rho_{\text{R}}(\y) \nonumber \\
   &\qquad  =2\pi ig\int d^4y
  \theta(y^-)\theta(y^-)G_0(x-y)\partial_y^i(W_y \rho_{\text{R}}(\y))
  \nonumber \\ & \qquad \qquad \qquad +4\pi ig \int_{y^-=0} dy^+d\y \int dz
 G_0(x-z)\delta(z^+)G_0(z-y)\theta(-y^+)W_z
    \partial^i_y \rho_{\text{R}}(\y), \label{sss} \end{align} where we used
    $W(y^+)=\theta(y^+)W + \theta(-y^+)$ since the support of
    $\rho_{\text{L}}$ (hence that of $A^-_{\mbox{\scriptsize{cl}}}$)
     is small in the $x^+$ direction.
 One can simplify the second term in (\ref{sss})
\begin{align} &  \int_{z^+=y^-=0} dy^+d\y G_0(z-y)\theta(-y^+)W_z
   \partial^i_y \rho_{\text{R}}(\y)\nonumber \\ & \qquad =\int d\y \frac{dp^-dp^+d\bm{p}}{(2\pi)^4}
   \frac{e^{-ip^+z^- +i\bm{p}(\z-\y)}}{2p^-p^+-\bm{p}^2+ip^+\epsilon}
   \frac{-i}{p^--i\epsilon}W_z
   \partial^i_y \rho_{\text{R}}(\y) \nonumber \\ & \qquad =\int d\y \frac{dp^+d\bm{p}}{(2\pi)^3}
   \frac{e^{-ip^+z^- +i\bm{p}(\z-\y)}}{-\bm{p}^2+i\epsilon}
   W_z
   \partial^i_y \rho_{\text{R}}(\y) \nonumber \\ & \qquad =\delta(z^-)
   W_z \frac{\partial^i_z}{\bm{\nabla}^2}
    \rho_{\text{R}}(\z)  \end{align}
 and the result for the produced gluon field is
 \begin{align} & \langle A^i(x) \rangle =2\pi ig\Bigl(\int d^4y
  \theta(y^-)\theta(y^+)G_0(x-y)\partial_y^i(W_y \rho_{\text{R}}(\y))
  \nonumber \\ &  \qquad \qquad \qquad +2 \int dz
 G_0(x-z)\delta(z^+)\delta(z^-)
   W_z \frac{\partial^i_z}{\bm{\nabla}^2}
    \rho_{\text{R}}(\z)\Bigr), \end{align} in agreement with \cite{tan}.

\section{Revisiting the Bremsstrahlung Hamiltonian}
\setcounter{equation}{0} In \cite{HIMST}, the dual partner of the
JIMWLK Hamiltonian--the Bremsstrahlung Hamiltonian
$H_{\text{BREM}}$--was derived. [See also an earlier derivation
 in \cite{KL1}.] However, its concrete usage was not clearly explained
there, nor was the source term specified.  In this section, we
complete the calculation of \cite{HIMST} by using the source terms
obtained in Section~3. As we shall see, this step is crucial for
the proper discussion of \emph{duality }\cite{KL2,HIMST} of the
evolution kernel. We  also show how our formulae of the
 $S$--matrix containing two eikonal phases (e.g., Eq.~(\ref{lc})) reduce to the
  CGC--type formula (see Eq.~(\ref{perio}) below)
 of the $S$--matrix
 containing only one eikonal phase.

We consider exactly the same situation as in Section 4.2, namely,
 scattering of a dilute right--mover ($\rho_{\text{R}} \sim 1$) and
a dense left--mover ($\rho_{\text{L}} \sim 1/g^2$) in the $A^+=0$
gauge. The forward $S$--matrix is (c.f., Eqs.~(\ref{lc}),
(\ref{31}))
\begin{align} S_Y= \int [D\rho_{\text{R}}(x^+)D\rho_{\text{L}}]
\int [DA^\mu] e^{iS_{\text{YM}}[A]-ig\rho_{\text{R}}(x^+) A^- +
iF^{-i}_{\text{cl}}
A_i(x^-=\infty)+iS_{\text{WZ}}[\rho_{\text{R}}(x^+)] }.
\label{iiii}
\end{align} We expand $A^\mu$ around a sum of
 solutions to the single source problem  \begin{align}
A^\mu_{\text{cl}} =B^i(x^+,x^-,\x)+ A^-(x^+,\x),
\end{align} where \begin{align} -D_i\partial^+ B^i=g\rho_{\text{R}}(x^+),
\qquad -\bm{\nabla}^2  A^-=g\rho_{\text{L}}. \label{yyu}
\end{align} Note that $F^{+-}_{\text{cl}}=0$. In
\cite{HIMST}, the current conservation law
$D^-_{\mbox{\scriptsize{cl}}}\rho_{\text{R}}=0$ was
 imposed. However, the first equation of Eq.~(\ref{yyu}) has a
 solution of the pure gauge type $F^{ij}_{\text{cl}}=0$
irrespective of the conservation law, and below we do not impose
it. After a shift $A^\mu =A^\mu_{\mbox{\scriptsize{cl}}}+a^\mu$,
Eq.~(\ref{iiii}) reads
\begin{align} & S_Y \approx \int
[D\rho_{\text{R}}(x^+)D\rho_{\text{L}}]\,
e^{iS_{\text{WZ}}[\rho_{\text{R}}(x^+)]} \nonumber
\\& \qquad  \times \int [Da^\mu]
\exp\left(-ig\rho_{\text{R}}  A^- + iD_\mu F^{\mu
i}_{\mbox{\scriptsize{cl}}}a_i+ \frac{i}{2}a^\mu G^{-1}_{\mu \nu}[
A^-]a^\nu + \cdots\right). \label{gau}
\end{align} Notice that the leading term is  given by
a single eikonal phase $e^{-\rho A^-}$ due to nontrivial
cancellation
 between  \begin{align}
S_{\text{YM}}[A_{\mbox{\scriptsize{cl}}}]=F^{+i}_{\mbox{\scriptsize{cl}}}F^{-i}_{\mbox{\scriptsize{cl}}}
\approx -\partial^+B^i D^i A^- =-D_i\partial^+B^i A^-
=g\rho_{\text{R}}  A^-, \end{align}
 and \begin{align} F^{-i}_{\mbox{\scriptsize{cl}}} B_i(x^+,\infty,\x)\approx -D_i A^-
\partial^+ B^i =
 A^-D_i\partial^+ B^i=-g\rho_{\text{R}} A^-, \label{hen}
\end{align} where we used $B^i(x^-)=\theta(x^-)B^i(\infty)$. Since $|B^i| \ll |
A^-|$, we have kept only terms linear in $B^i$. It is crucial to
 specify the correct source terms to see this cancellation.

 Performing Gaussian integration in Eq.~(\ref{gau}), and
reinstating  the weight functions, we obtain
\begin{align} & S_{Y}=\int [D\rho_{\text{R}}(x^+)D\rho_{\text{L}}]
Z_{Y-Y_0}[\rho_{\text{R}}]Z_{Y_0}[\rho_{\text{L}}]
 \nonumber \\ & \qquad \qquad \times \exp\left(iS_{\text{WZ}}[\rho_{\text{R}}(x^+)]-ig\rho_{\text{R}}
 A^- +i\delta Y S_{\mbox{\scriptsize{BREM}}}[\rho_{\text{R}}(\pm \infty),\widetilde{W}]\right), \label{nin}
\end{align} where $\delta Y$ is the rapidity interval carried by
the $a^\mu$ field. As indicated, the right--mover carries rapidity
$Y-Y_0$ and the left--mover, $Y_0$. $S_{\mbox{\scriptsize{BREM}}}$
is an effective action describing gluon Bremsstrahlung whose
explicit form can be found in \cite{HIMST}. It contains charges
$\rho_{\text{R}}$ at $x^+=\pm \infty$ and the adjoint Wilson line
$\widetilde{W}$. From Eq.~(\ref{nin}) one can deduce the evolution
equation
\begin{align} \frac{\partial S_Y}{\partial Y} =\int
[D\rho_{\text{R}}(x^+)D\rho_{\text{L}}]
Z_{Y-Y_0}[\rho_{\text{R}}]Z_{Y_0}[\rho_{\text{L}}]
 \, iS_{\mbox{\scriptsize{BREM}}}[\rho_{\text{R}}(\pm \infty),\widetilde{W}]\exp\left(iS_{\text{WZ}}[\rho_{\text{R}}(x^+)]-ig\rho_{\text{R}}
 A^-\right). \label{yaki}
\end{align} There are two ways to proceed from Eq.~(\ref{yaki}):
\\
 (i) The eikonal factor
$e^{-ig\rho_{\text{R}}
 A^-}$ allows one to  replace $\rho_{\text{R}}(\pm \infty) \to i\delta /\delta
A^-(\pm \infty)$ in $S_{\mbox{\scriptsize{BREM}}}$ to convert the
action into an operator, the \emph{JIMWLK} Hamiltonian
\begin{align}
iS_{\mbox{\scriptsize{BREM}}}[\rho_{\pm\infty},e^{iA^-}] \to
-H_{\mbox{\scriptsize{JIMWLK}}}\Bigl[i\frac{\delta}{\delta
A^-_{\pm\infty}},e^{iA^-}\Bigr]. \end{align}
 After this is done, one can use the path integral formula Eq.~(\ref{form}) in the backward
direction and get
\begin{align} \frac{\partial S_Y}{\partial Y} =-\int D\rho_{\text{L}}
Z_{Y_0}[\rho_{\text{L}}]
 H_{\mbox{\scriptsize{JIMWLK}}}
 \Big[\frac{\delta}{\delta A^-_{\pm \infty}},\widetilde{W}(A^-(x^+))\Big]\langle \prod_i
 W(\x_i)\rangle_{Y-Y_0}, \label{ae}
\end{align} with \begin{align} & S_Y=\int D\rho_{\text{L}} Z_{Y_0}[\rho_{\text{L}}]
\, \langle \prod_i
 W(\x_i)\rangle_{Y-Y_0}, \nonumber \\ & \langle \prod_i
 W(\x_i)\rangle \equiv \int D\rho_{\text{R}} Z[\rho_{\text{R}}]\prod_i
 W(\x_i). \label{perio} \end{align}  The differential operator $H_{\text{JIMWLK}}$
 can act on either (left or right) direction through the integration by parts.
 When it acts on $Z[\rho_{\text{L}}]$, Eq.~(\ref{ae})  describes the evolution of
 the left--moving Color Glass Condensate, namely, the JIMWLK equation.
  When it acts on the right, Eq.~(\ref{ae}) describes the evolution of  Wilson lines,
   the Balitsky equation. In the latter case it may be more natural to put
   the energy ($Y-Y_0$)
dependence in the slope of the Wilson lines as originally done in
\cite{B}.

(ii)  Returning to Eq.~(\ref{yaki}) this time we replace $A^-(x^+)
\to i\delta/\delta \rho_{\text{R}}(x^+)$ in the adjoint Wilson
line $\widetilde{W}$ in $S_{\mbox{\scriptsize{BREM}}}$ to obtain
the \emph{Bremsstrahlung} Hamiltonian \begin{align}
iS_{\mbox{\scriptsize{BREM}}}[\rho_{\pm\infty},e^{iA^-}]
 \to -H_{\mbox{\scriptsize{BREM}}}[\rho_{\pm\infty},e^{-\frac{\delta}{\delta
 \rho}}].\end{align}
  After this is done,
$H_{\mbox{\scriptsize{BREM}}}$ has no reference to the left--mover
anymore,
 so we can drop the averaging over $Z[\rho_{\text{L}}]$
\begin{align} & \frac{\partial}{\partial Y}\langle \prod_i
 W(\x_i)\rangle=-\int D\rho_{\text{R}}(x^+)
 Z[\rho_{\text{R}}]e^{iS_{\text{WZ}}[\rho_{\text{R}}(x^+)]} \nonumber \\
 & \qquad \qquad \times H_{\mbox{\scriptsize{BREM}}}\Bigl[\rho_{\text{R}}(\pm \infty),
 \widetilde{W}\bigl(\frac{\delta}{\delta \rho_{\text{R}}(x^+)}\bigr)\Bigr]
 \exp\left(-ig\rho_{\text{R}}
 A^-\right).
 \end{align} Functionally differentiating both sides with respect to $A^-(x^+)$
 arbitrary times and setting $A^-=0$, we obtain an equation for the correlator of
 non--commutative charges \begin{align}
 & \frac{\partial}{\partial Y} \langle \hat{\rho_{\text{R}}}^a \hat{\rho_{\text{R}}}^b
 \hat{\rho_{\text{R}}}^c \cdots \rangle =
 -\int D\rho_{\text{R}}(x^+) e^{iS_{\text{WZ}}[\rho_{\text{R}}(x^+)]} Z[\rho_{\text{R}}] \nonumber \\
  & \qquad \times H_{\mbox{\scriptsize{BREM}}}\Bigl[\rho_{\text{R}}(\pm \infty),
 \widetilde{W}\bigl(\frac{\delta}{\delta
 \rho_{\text{R}}(x^+)}\bigr)\Bigr]\rho_{\text{R}}^{a}(x^+_1)
 \rho_{\text{R}}^{b}(x^+_2)\rho_{\text{R}}^c(x^+_3) \cdots, \label{o} \end{align}
 where the `ordering variables' $x^+_i$ satisfies the constraint
 $x^+_1 > x^+_2 > x^+_3 > \cdots$ but otherwise arbitrary.
 Eq.~(\ref{o}) coincides with the equations derived in
 \cite{KL1}. Our derivation unambiguously shows that the correct ordering variable
  is $x^+$, and not $x^-$ as claimed in \cite{KL1,KL5}.

 The above calculations nicely exemplify the duality property of the evolution
 kernel \cite{KL2,HIMST}. The effective action $S_{\mbox{\scriptsize{BREM}}}$
 can simultaneously
 describe the gluon saturation of  the left--mover (Eq.~(\ref{ae})) and the
 gluon Bremsstrahlung of the right--mover (Eq.~(\ref{o})) in  asymmetric scattering.

\section{Conclusion}
In this paper we considered scattering of two saturated hadrons in
the functional integral framework. We have constructed the source
term for each of the hadrons including the case of the light--cone
gauge. Our formulation is applicable to colliding hadrons at any
stage of their high energy evolution, from the dilute regime
$\rho\sim 1$ to the dense regime $\rho \sim 1/g^2$. Once one knows
the proper source term, issues such as gluon production and the
evolution of the $S$--matrix become more transparent and
straightforward to deal with as demonstrated for the $pA$
collision. We hope to address the problem of the $AA$ collisions
in future work.

\section*{Acknowledgements}
I am grateful  to Edmond Iancu for many discussions and
encouragement. I also thank Ian Balitsky, Kenji Fukushima,  Lev
Lipatov, Larry McLerran and Feng Yuan for discussions, and the
anonymous referee for constructive remarks which helped me to
improve the presentation of Sections~2--3. This work is supported
by RIKEN, Brookhaven National Laboratory and the U.~S.~Department
of Energy [Contract No. DE-AC02-98CH10886].

\appendix

\section{Path integral of a spin}

In this appendix we give a very quick review of the standard path
integral procedure for a spin operator. Rigorous treatment can be
found in the literature  \cite{wz}. For the sake of simplicity, we
confine ourselves to the SU(2) spin.

For an ordinary  quantum
 mechanical system endowed with a Hamiltonian $H[\hat{q},\hat{p}]$
 ($\hat{q}$ is the canonical coordinate and $\hat{p}$ is its conjugate
 momentum), the path integral representation of the transition amplitude
  is elementary, and is given by  \begin{align} \langle
q_f|\, e^{-iTH[\hat{p},\hat{q}]}|q_i
\rangle=\int_{q(0)=q_i}^{q(T)=q_f} [Dq(t)Dp(t)]\exp \left( i
\int_0^{T} dt \bigl( p(t)\dot{q}(t) - H[p(t),q(t)] \bigr) \right).
\label{first}
 \end{align}
If the Hamiltonian contains spin operators, the path integral
becomes less straightforward. As the simplest example, consider a
non--relativistic particle with the spin operators $\hat{S}_a$
($a=x,y,z$) in the $J$-representation $(J=1/2,1,3/2, \cdots$) of
SU(2) placed in a constant magnetic field $\vec{B}$ pointing to
the $z$-direction. The interaction Hamiltonian is
\begin{align} H=-\vec{\hat{S}}\cdot \vec{B}=-\hat{S}_zB.
\label{hami}
\end{align}
   The spin operators satisfy the usual commutation
relation $[\hat{S}_a,\hat{S}_b]=i\epsilon_{abc}\hat{S}_c$. Our
goal is to write down a path integral formula
 similar to
 Eq.~(\ref{first}) for the Hamiltonian Eq.~(\ref{hami}). We must begin with
 finding out
 what are $q$ and $p$ in this case.

In the path integral representation, a quantum operator $\hat{q}$
has its classical counterpart $q(t)$. To the spin operator
 corresponds a rotating vector on a sphere \begin{align}
\vec{S}(t)=J\vec{n}(t)=J(\sin \theta \cos
 \phi, \sin \theta \sin \phi, \cos \theta), \label{spin} \end{align} where
  $\vec{n}$ is a unit vector. Though $\vec{S}$ has
 three components, there are only two independent variables, $\theta$ and
 $\phi$. In fact, $\phi$ plays the role of the canonical
 coordinate,  and the conjugate momentum is $\cos \theta$
 \begin{align} q(t)=\phi(t), \qquad  p(t)=J\cos \theta(t). \end{align} Indeed,
  if one imposes the Poisson bracket \begin{align}
 \{\phi, J\cos\theta\}_{\mbox{\scriptsize{P.B.}}}=1, \quad \quad \{\theta,
 \phi\}_{\mbox{\scriptsize{P.B.}}}=\frac{1}{J\sin\theta}, \label{po} \end{align}
   from Eq.~(\ref{spin}) one can easily
 derive the expected Poisson brackets among the components of the
  vector $\vec{S}$ \footnote{As usual, this
 differs from the quantum commutation relation $[\hat{S}^a,
 \hat{S}^b]=i\epsilon^{abc}\hat{S}^c$ by a factor $i$.}
 \begin{align} \{S^a,
 S^b\}_{\mbox{\scriptsize{P.B.}}}=\epsilon^{abc}S^c. \label{com}\end{align}

 With the canonical coordinate and
 momentum at hand, we naively insert them into Eq.~(\ref{first}) and
  obtain a path integral formula
 for the Hamiltonian Eq.~(\ref{hami})
  \begin{align} \langle
\phi_f|e^{-iTH[\vec{S}]}|\phi_i
\rangle=\int_{\phi(0)=\phi_i}^{\phi(T)=\phi_f}
[D\phi(t)D\cos\theta(t)]\exp \left( i \int_0^{T} dt \bigl(
 J\cos\theta \dot{\phi} + BJ\cos\theta  \bigr) \right).
 \label{p} \end{align}  It turns out that Eq.~(\ref{p}) is (more or less) the correct
 formula, though the
 actual proof \cite{wz} requires careful manipulation of
 integrals.
 One starts with the  discretized  form
 of the right hand side
 of Eq.~(\ref{p})
 \begin{align}
\sum_{n=-\infty}^{\infty} \lim_{N\to \infty}
\frac{1}{(2\pi)^N}\int_{\phi_i}^{\phi'_f}\prod_{k=1}^{N} D\cos
\theta_k \prod_{k=1}^{N-1} D\phi_k
\exp{\left[i\sum_{k=1}^N\left(\epsilon \bigl(J\cos\theta_k+\gamma
\bigr)\frac{\phi_k-\phi_{k-1}}{\epsilon}+\epsilon BJ\cos
\theta_k\right)\right]},
 \label{pp} \end{align} where $T=N\epsilon$, $\phi'_f\equiv
\phi_f+2\pi n=\phi_N$, $\phi_i=\phi_0$, and shows that it reduces
to the left hand side  \begin{align}
 \langle
\phi_f|\, e^{-iTH[\vec{S}]}|\phi_i \rangle= \frac{1}{2\pi}
\sum_{S_z=-J}^{J}e^{i(\phi_f-\phi_i+BT)S_z}.
\end{align}
 A constant $\gamma$ is
 needed in Eq.~(\ref{pp}) for a technical reason.
 For  half--integer $J$, one can choose
 $\gamma=\frac{1}{2}$.

The kinetic term $\int p \dot{q} \sim \int \cos \theta\cdot
\dot{\phi} $ in Eq.~(\ref{p}) is what we call `the Wess--Zumino
term'.  To obtain the formula for a Wilson line, Eq.~(\ref{form})
in the main text, one just needs to make the following
replacements in Eq.~(\ref{p})  \begin{align} t \to x^+, \qquad B_a
\to gA^-_a(x^+), \qquad \hat{S}^a \to t^a, \qquad S^a(t) \to -
\rho^a(x^+), \label{just}
\end{align} and use a sloppy notation for the measure $D\phi(t)D\cos \theta(t) \to
D\rho(x^+)$.  In practice, one is interested in  the transition
amplitude between states with definite color indices $W_{ij}(\x)$,
rather than definite values of the conjugate variable $\phi$. In
this case one has to project onto these color states by
integrating over the initial
   $\phi_i$ and final
  $\phi_f$ angles: $\int_0^{2\pi} d\phi_{i,f} e^{iS_z
  \phi_{i,f}}$.

 The Wess--Zumino term has a simple representation in
terms of a matrix  \begin{align} S_{\text{WZ}}=J\int dx^+
\cos\theta \dot{\phi} = iJ\int dx^+ \tr [\tau^3 M
\partial_{+}M^\dagger], \label{mat} \end{align} where $\tau^3$ is the third Pauli matrix
 and  \begin{align} M=\left(%
\begin{array}{cc}
  \cos\frac{\theta}{2}e^{i\frac{\phi}{2}}, & \sin\frac{\theta}{2}e^{-i\frac{\phi}{2}} \\
  -\sin \frac{\theta}{2}e^{i\frac{\phi}{2}}, & \cos \frac{\theta}{2}e^{-i\frac{\phi}{2}} \\
 \end{array}%
\right).
\end{align} Furthermore, $M$ is related to $\rho$ via
\begin{align} \rho^a=-\frac{J}{2} \tr [\tau^3 M \tau^a M^\dagger].
\label{related}
\end{align}
 Eqs.~(\ref{mat}) and (\ref{related}) are the forms used in \cite{hatta}
 (there the letter $S$ was used instead of $M$).
 Since Eq.~(\ref{p}) (or Eq.~(\ref{form})) is the genuine path
 integral for the spin, the non--commutativity of $\hat{S}^a$ (or
 $t^a$) is guaranteed to be preserved. Yet,  in the path
 integral representation, of course one can treat $S^a(t)$ (or $\rho(x^+)$) as
 c--numbers. For instance, the following identity
 holds
  \begin{align} \langle f|\,  t^{a_1}t^{a_2}t^{a_3}
\cdots t^{a_n} |i \rangle=\int [D\rho(x^+)] \,
e^{iS_{\text{WZ}}[\rho(x^+)]}\rho^{a_1}(x^+_{a_1})
\rho^{a_2}(x^+_{a_2})\rho^{a_3}(x^+_{a_3})\cdots
 \rho^{a_n}(x^+_{a_n}),
 \label{abo} \end{align} where $x^+$'s are constrained such that
 $x^+_{a_1}>x^+_{a_2}>x^+_{a_3}> \cdots >
 x^+_{a_n}$, but otherwise arbitrary (only the ordering matters). If one changes
 the order of  matrices, say,
 $t^{a_2}$ and $t^{a_3}$, one correctly obtains the commutator
 term $if^{a_2 a_3 c}\rho^c$ on the right hand side.

\end{document}